\pdfoutput=1

\def\ttitle{Phase space analysis of two-wavelength interferometry}
\def\kkeywords{
  interferometry,
  multi-wavelength interferometry,
  two-wavelength interferometry,
  algorithm,
  phase space,
  error analysis,
}

\documentclass[a4paper,9pt,twocolumn]{article}

\usepackage{ifthen}
\usepackage{charter}
\usepackage{cancel}
\usepackage{makeidx}
\usepackage{amsthm}
\usepackage{amssymb}
\usepackage{amsfonts}
\usepackage{graphicx}
\usepackage{MnSymbol}
\usepackage{float}
\usepackage[numbers,sort&compress]{natbib}
\usepackage[pdfpagelabels=true]{hyperref}
\usepackage{hypernat}
\usepackage[letterpaper, margin=1in]{geometry}

\usepackage{tikz}
\usepackage{gnuplot-lua-tikz}
\usepackage{IEEEtrantools}

\usepackage{array}
\usepackage{amsmath}
\usepackage{tabularx}
\usepackage{yfonts}
\usepackage{fancyhdr}
\usepackage{authblk}
\usepackage[nolist]{acronym}
\usepackage{./sty/shortcuts} 
\usepackage{eso-pic}
\usepackage{color}
\usepackage{type1cm}
\usepackage{import}      
\usepackage{mathtools, nccmath}
\usepackage{gensymb}
\usepackage{abstract}
\usepackage{sectsty}

\sectionfont{\fontsize{12}{15}\selectfont}

\hypersetup{
    naturalnames=true,
    colorlinks=true,
    linkcolor=blue,
    pdfpagemode=UseNone,
    pdfstartview=FitH,
    pdftitle={\ttitle},
    pdfauthor=Air Force Research Laboratory,
    pdfsubject={Phase Space Analysis of Two-Wavelength Interferometry},
    pdfkeywords={\kkeywords},
    debug,
}

\newcolumntype{L}[1]{>{\raggedright\let\newline\\\arraybackslash}m{#1}}
\newcolumntype{C}[1]{>{\centering\let\newline\\\arraybackslash}m{#1}}
\newcolumntype{R}[1]{>{\raggedleft\let\newline\\\arraybackslash}m{#1}}

\makeatletter
\newcommand{\vast}{\bBigg@{4}}
\newcommand{\Vast}{\bBigg@{5}}
\makeatother

\begin{acronym}
  \acro{OPD}{optical path difference}
  \acro{UR}{unambiguous range}
  \acro{HC}{Houairi \& Cassaing}
\end{acronym}

\title{\ttitle}

\author[1,*]{Robert H. Leonard}
\author[2]{Spencer E. Olson}
\affil[1]{Space Dynamics Laboratory, Quantum Sensing \& Timing, North Logan, UT 84341, USA}
\affil[2]{Air Force Research Laboratory, Kirtland Air Force Base, NM 87117, USA}

\affil[*]{robert.leonard@sdl.usu.edu}

\newcommand{\distA}[1]{%
  Approved for public release; distribution is unlimited.  Public Affairs %
  release approval %
  #1.
}

\begin{document}

\twocolumn[
  \maketitle
  \thispagestyle{fancy}
  \begin{onecolabstract}
  \begin{minipage}{1.0\linewidth}
    Multiple wavelength phase shifting interferometry is widely used to extend the
    \ac{UR} beyond that of a single wavelength.  Towards this end,
    many algorithms have been developed to calculate the \ac{OPD}
    from the phase measurements of multiple wavelengths.  These algorithms fail when
    phase error exceeds a specific threshold. In this paper, we examine this failure
    condition.  We introduce a ``phase-space'' view of
    multi-wavelength algorithms and demonstrate how this view may be used to
    understand an algorithm's robustness to phase measurement error.  In particular,
    we show that the robustness of the synthetic wavelength algorithm deteriorates
    near the edges of its \ac{UR}.  We show that the robustness of de~Groot's
    extended range algorithm~\cite{deGroot_1994} depends on
    both wavelength and \ac{OPD} in a non-trivial manner.  Further, we demonstrate
    that the algorithm developed by \ac{HC}~\cite{Houairi_2009} results in uniform robustness across the entire
    \ac{UR}.  Finally, we explore the effect that wavelength error has on the
    robustness of the \ac{HC} algorithm.
  \end{minipage}
  \end{onecolabstract}
]

\acresetall

\section{Introduction}
Determination of the absolute phase is important in applications such
as interferometric synthetic aperture
radar~\cite{Chiaradia_1996,Veneziani_2003,Xu_2016},
strain/stress analysis~\cite{Siebert_2004},
and atom interferometry~\cite{Yankelev_2020}.  Phase measurements are
also critical in order to characterize a surface height via optical
profilometry~\cite{Wyant_1984,Ribbens_1974}, which can be used with a single
wavelength or a more complex multiple-wavelength configuration.

Single-wavelength phase shifting interferometry measures the phase
difference modulo $2\pi$, resulting in an $n \lambda$ ambiguity in the \ac{OPD},
where $n$ is an integer and $\lambda$ is the wavelength used.
Consequently, the \ac{OPD} may be unambiguously resolved when when restricted to the
range $\pm\lambda/2$; we refer to such a range as the \ac{UR}, and denote the
length of the range as $\abs{\ac{UR}}$.
Comparison of phase measurements from multiple wavelengths allows the \ac{UR} to
be extended beyond that of a single wavelength.  Towards this end, many
algorithms have been developed which incorporate phase measurements from
multiple wavelengths to determine the absolute phase with a larger
\ac{UR}~\cite{Polhemus_1973,deGroot_1994,Houairi_2009,Falaggis_2008,Lofdahl_2001}.
These algorithms will fail when the phase error exceeds a specific threshold.

In this paper we explore the conditions under which multi-wavelength algorithms
are valid.  In Sec.~\ref{sec:max_ur} we provide a brief derivation of the
maximum \ac{UR} achievable for an algorithm which uses two phase measurements.
In Sec.~\ref{sec:phase_space} we introduce and characterize the phase space of
multi-wavelength interferometry.  In Sec.~\ref{sec:phase_space_error} we build
on our understanding of the phase space to describe the effect that phase
measurement error has on the accuracy of the measured \ac{OPD} as well as the
robustness of the algorithm.  In Sec.~\ref{sec:synthetic} we show that the
synthetic wavelength algorithm, where data from multiple wavelengths are
combined together to create a single data set of larger synthetic wavelength
$\Lambda$, creates an uneven partitioning of the phase
space.  This uneven partitioning results in a robustness which decreases when
the \ac{OPD} is within $\lambda_{i}/2$ of $\pm\Lambda/2$, where $\lambda_i$ is
any of the individual real wavelengths.  We show that this decrease in
robustness is the result of an overlooked constraint of the algorithm. In
Sec.~\ref{sec:de_groot}, we examine an algorithm created by
de~Groot~\cite{deGroot_1994}.  While de~Groot's algorithm correctly calculates the
\ac{OPD} in the absence of phase measurement error, we find that the algorithm
exhibits complicated behavior when phase error is present.  An explanation for
this behavior is presented.  We show that the robustness of de~Groot's algorithm
will generally vary with \ac{OPD}. The condition under which de~Groot's algorithm
achieves maximum robustness is described.  In Sec.~\ref{sec:hc_algorithm}, we
demonstrate that the Houairi \& Cassaing algorithm evenly partitions the phase
space, resulting in uniform robustness over nearly the entire \ac{UR}.  In
Sec.~\ref{sec:wavelength_error} we examine the effect of wavelength error on the
\ac{HC} algorithm.

\section{Maximum Unambiguous Range}\label{sec:max_ur}
We define the maximum \ac{UR} as the largest range of \ac{OPD} such that any
two \acp{OPD} within this range are distinguishable when measurement error is
absent. For an algorithm which uses a pair of phase measurements from two
different wavelengths as input, two \acp{OPD} are indistinguishable when they
result in the same measured phase pair $(\phi_a,\phi_b)$ where $a$ and $b$
denote data from the different wavelengths. Therefore, the minimum distance
between two \acp{OPD} which result in the same phase pair will equal the maximum
\ac{UR}.  We will denote the \ac{OPD} as $d$.  To simplify the analysis, we will
assume  $\lambda_a > \lambda_b$ throughout.

Consider the measured phase pair $(0,0)$.  A phase measurement will equal zero
when $d$ equals an integer multiple of $\lambda$.  Therefore, the phases pass
through $(0,0)$ when $d = n_a \lambda_a = n_b \lambda_b$ where
$n_a,n_b \in \mathbb{Z}$. This Diophantine equation has a trivial solution at
$d=0$. Following the notation used by
Houairi and Cassaing~\cite{Houairi_2009},
we denote the next smallest integer pair which satisfies this equation as $p,q$.
Therefore, the length of the maximum $\abs{\ac{UR}}$ may be written as

\begin{equation}
  \label{eq:UR_phase_condition1}
  \abs{\ac{UR}_{\rm max}} = p \lambda_b = q \lambda_a
\end{equation}

\noindent where $p,q$ are the co-prime natural numbers which satisfy the
equation on the right side of Eq.~\ref{eq:UR_phase_condition1}.
This result has been previously noted~\cite{Falaggis_2008,Brug_1998}.

Note that Eq.~\ref{eq:UR_phase_condition1} is only satisfied when $\lambda_a$
and $\lambda_b$ are commensurate.  When $\lambda_a$ and $\lambda_b$ are
incommensurate, the measured phase pairs never repeat, resulting in an infinite
\ac{UR}.  In practice, the maximum desirable \ac{UR} will be constrained by measurement
error~\cite{Houairi_2009}.

\section{Phase-Space Representation}\label{sec:phase_space}
For two-wavelength interferometry with wavelengths $\lambda_a$ and $\lambda_b$,
the measured phases in the absence of error are given by
\begin{equation}
  \label{eq:phase}
  \phi_i = \frac{2\pi\, d}{\lambda_i} \mod 2\pi
\end{equation}
where $i \in \{a,b\}$, and $\mod$ is defined by the formula
\begin{equation}
  \label{eq:modulo}
  a \mod b \coloneqq a - b \, \round{ \frac{a}{b} }
\end{equation}
and where $\round{x}$ denotes the nearest integer value to $x$.

\begin{figure}[htb!]
  \vspace*{0.2cm}
  \centering
  \includegraphics[width=\columnwidth]{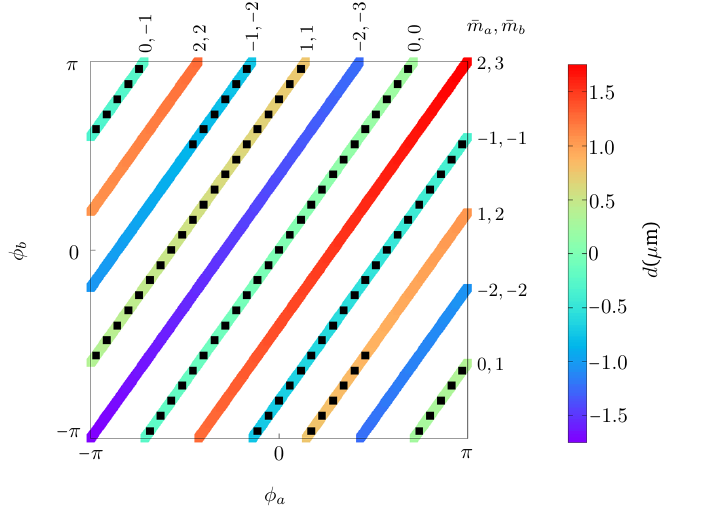}
  \caption[Two Wavelength Interferometry Phase Space]{
    Ideal phase measurements for $\lambda_a = 700~\nm$ and
    $\lambda_b = 500~\nm$ over: the entire  $3.5~\um$ \ac{UR} (solid);
    the \ac{UR} of the synthetic wavelength algorithm (square dots).  The
    \ac{OPD} ($d$) is represented by the color.  At $d=0~\um$ the phase is
    $(0,0)$.  As $d$ increases the phases moves upwards along the $0,0$ line.
    When $\phi_b=\pi$, the phases jump to the bottom of the $0,1$ line.
    Starting from -$\ac{UR}_{\rm max}/2$, and increasing in $d$, the ideal phases jump
    between the lines in the following order: $(-2-3)$, $(-2-2)$, $(-1,-2)$,
    $(-1,-1)$, $(0,-1)$, $(0,0)$, $(0,1)$, $(1,1)$, $(1,2)$, $(2,2)$, $(2,3)$.
  }
  \label{fig:phase_space}
\end{figure}

Consider a graph in which $\phi_a$ and $\phi_b$ are plotted on the $x$ and $y$
axes respectively.  In this representation, all measured phases fall within the
space $(\phi_a,\phi_b) \in [-\pi,\pi) \times [-\pi,\pi)$; we call this
\textit{phase space}.  In Fig.~\ref{fig:phase_space}, we show ideal phase measurements
plotted in phase space over the entire \ac{UR}. This representation is easily
constructed by parametrically plotting the phases using Eq.~\ref{eq:phase}.

Following the notation introduced in Ref.~\cite{Houairi_2009}, we write the
\ac{OPD} ($d$) as follows

\begin{equation}
  \label{eq:h_opd}
  d = \lambda_a (\bar{m}_a+\dot{m}_a) = \lambda_b (\bar{m}_b+\dot{m}_b)
\end{equation}

\noindent where $\bar{m} \in \mathbb{Z}$ and $\dot{m} \in [-0.5,0.5]$.  The
value of $\dot{m}$ is related to the measured phase by

\begin{equation}
  \label{eq:m_dot}
\dot{m}_i = \frac{\phi_i}{2 \pi} .
\end{equation}

As shown in Fig.~\ref{fig:phase_space}, as $d$ increases from zero which is
found at $\left(\phi_a, \phi_b\right) = (0,0)$, the phases follow a line with slope
$r_{\lambda} = \lambda_a/\lambda_b$ in the phase space.  When $\phi_a$ or
$\phi_b$ reach $\pm \pi$, the phase ``wraps back'' to $\mp\pi$, creating a new line.
This occurs when $\dot{m}_i = \pm0.5$. The new line corresponds to the incremented
value of $\bar{m}_i$.  In this way, ideal phase measurements form a set of
parallel lines where each line is uniquely associated with an integer pair
$(\bar{m}_a,\bar{m}_b)$.  When plotted over the entire \ac{UR}, each line is
displaced from the adjacent line by an equal amount.  The displacement between
adjacent lines is

\begin{equation}
    \label{eq:phase_offset}
    \Delta \phi_i = \frac{2\pi \lambda_i}{\abs{\ac{UR}_{\rm max}}} .
\end{equation}

\section{Phase-Space View of Phase Errors}\label{sec:phase_space_error}

Building on our knowledge of the phase space, we can imagine a two-wavelength
interferometry algorithm which achieves the maximal \ac{UR}.  When error is present,
measured phases deviate from the lines shown in Fig.~\ref{fig:phase_space}.
Because each line is associated with a unique integer pair
$(\bar{m}_a , \bar{m}_b)$, we can determine the values of $\bar{m}_i$ by
identifying the line closest to the measured phase pair $(\phi_a,\phi_b)$. Once the values of
$\bar{m}_i$ are known, the \ac{OPD} ($d$) is determined using to Eqs.~\ref{eq:h_opd} and
\ref{eq:m_dot}.  Brug and Klaver~\cite{Brug_1998} developed an algorithm similar
to this using a lookup table for all phase values in the phase space.

Phase measurement error effects interferometry algorithms in two ways: (1) phase
error may result in a error in the calculated value of $\dot{m}_i$ (a continuous
variable); (2) phase error may result
in the incorrect determination of $\bar{m}_i$ (an integer variable).  Because an incorrect
determination of $\bar{m}_i$ will result in a large error in the calculated
\ac{OPD},
we classify \textbf{(2)} as an algorithm failure and \textbf{(1)} as typical
measurement error.  It is helpful to think of the phase errors as a
displacement vector in the phase space,
$\vec{\delta \phi} = \langle \delta \phi_a, \delta \phi_b \rangle$ where
$\delta \phi_i$ is the error in $\phi_i$.
When possible, it is helpful to
decompose the phase error into components corresponding to these different
effects.  To this end, another convenient coordinate system
for error decomposition is parallel and perpendicular to the ideal phase
lines (e.g as shown in Fig.~\ref{fig:phase_space}), denoted by $\vec{\delta \phi} = \langle \delta \phi_\perp, \delta
\phi_\parallel \rangle$.

An algorithm will incorrectly determine $\bar{m}_i$ when phase error displaces
the phase from its ideal value so that the measured phase is closer to a line
corresponding to incorrect values of $\bar{m}_i$.  Consequently, $\delta\phi_\perp$ is
responsible for algorithm failure, whereas $\delta\phi_\parallel$ is identified
as simple measurement error.

Any two-wavelength algorithm which uses Eq.~\ref{eq:h_opd} will result in two
\ac{OPD} calculations.  In general, when both calculated values for \ac{OPD} are combined into a
weighted average, the resulting error in the \ac{OPD} will depend on the phase
error in a non-trivial way. For the special case in which both phase
measurements have equal uncertainty, the weighted average of the two
\ac{OPD} results will correspond to a point exactly on the ideal phase line which is closest
to the measured phase.  In this case, $\delta\phi_\parallel$
is solely responsible for error in the calculated \ac{OPD}.  We will
assume that both \ac{OPD} results are combined into a weighted average for the
remainder of this paper.  When the phase uncertainties are the same for both
wavelengths, $\delta \phi_{\parallel}$ is related to the error in $d$ by
\begin{equation}
  \label{eq:parallel_error_relation2}
  \delta d = \frac{\delta \phi_{\parallel}}{2 \pi} \frac{\lambda_a \lambda_b}{\sqrt{\lambda_a^2 + \lambda_b^2}}
  .
\end{equation}
$\langle \phi_a, \phi_b \rangle$ is related to
$\langle \phi_{\parallel}, \phi_{\perp} \rangle$ by the rotational transformation
\begin{equation}
  \label{eq:basis_transformation}
  \langle \phi_{\parallel}, \phi_{\perp} \rangle^{\mbox{T}} =
    \boldsymbol{\mathcal{R}}_{-\theta} \langle \phi_a, \phi_b \rangle^{\mbox{T}}
\end{equation}
where $\theta$ is given by
\begin{equation}
  \label{eq:transformation_angle}
  \theta = \tan^{-1}\left(\frac{\lambda_a}{\lambda_b}\right)
\end{equation}
and where $\boldsymbol{\mathcal{R}}_{-\theta}$ is the typical rotation matrix about an
angle $-\theta$.

Recall that an algorithm is deemed to have failed when the values of $\bar{m}_i$ are
determined incorrectly. We define robustness, $R$, as the probability that the
algorithm will succeed.   Robustness may be calculated by integrating the
probability distribution function for the measured phase values over the range
of phase values for which the algorithm returns the correct values of
$\bar{m}_i$.  For the remainder of this paper, we assume that measured
phases deviate from ideal values according to a normal distribution with
standard deviation equal to the measurement uncertainty, $\sigma_i$.  As we
see in the next section, the robustness may depend on the \ac{OPD}.

Note that any algorithm which achieves the maximum \ac{UR} will pass through
the $(\pm \pi, \pm \pi)$ corners of the phase space when $d = \pm{\rm \ac{UR}}_{\rm max}/2$.
Because measured phases are constrained to fall within $[-\pi,\pi)$,
discontinuities in the measured phases arise near the $(\pm \pi, \pm \pi)$
corners (Fig.~\ref{fig:de_groot_discontiuities}a can be used to visualize this).
For instance: when
$d = -\ac{UR}_{\rm max}/2$, the phases should be $(-\pi,-\pi)$.  A meaurement
error which would normally result in a small negative phase error, now results
in a phase error of nearly $+2 \pi$ (as measurement error can cause $\phi_a$ to
wrap to the right of Fig.~\ref{fig:de_groot_discontiuities}a or
$\phi_b$ to wrap to the top of Fig.~\ref{fig:de_groot_discontiuities}a).  As a
result, the robustness of an algorithm
which achieves the maximum \ac{UR} will deteriorate when the \ac{OPD} is within
measurement uncertainty of $\pm \abs{\ac{UR}_{\rm max}}/2$.  Specifically, the
robustness will approach $1/4$ as $d \to \pm \abs{\ac{UR}_{\rm max}}/2$.  An
algorithm will avoid this failure mode when both phase errors satisfy the
constraint
\begin{equation}
  \label{eq:HC_constraint3}
  \left \lvert d_0 + \lambda_i \frac{\delta \phi_i}{2 \pi} \right \rvert <
  \frac{\abs{\ac{UR}_{\rm max}}}{2}
  .
\end{equation}

\section{Synthetic Wavelength Algorithm}\label{sec:synthetic}

The synthetic wavelength algorithm was first introduced by J.~C. Wyant~\cite{Wyant71}.
The algorithm uses phase measurements from two wavelengths, $\lambda_a$ and
$\lambda_b$, to calculate the phase that would be produced by a larger synthetic
wavelength, $\Lambda$, which is defined as
\begin{equation}
  \label{eq:synthetic_wavelength}
  \Lambda = \frac{\lambda_a \, \lambda_b}{\left|\lambda_a - \lambda_b\right|}.
\end{equation}
As a result, the synthetic wavelength algorithm has an \ac{UR} of
$\pm \Lambda/2$.  A major benefit of phase measurements from an effectively
longer wavelength is to simplify the phase-unwrapping by reducing the many
$2\pi$ phase ambiguities.  Additionally, a synthetic wavelength measurement
produced from shorter wavelengths takes advantage of much easier optical
elements, detectors, and cameras as compared to available resources for the
longer wavelengths.

For most choices of wavelengths, the \ac{UR} of the synthetic wavelength algorithm
is less than the maximum \ac{UR} that could be achieved for same choice of
wavelengths (as described in Sec.~\ref{sec:max_ur}).  When ideal phase
measurements are plotted over the \ac{UR} of the synthetic wavelength algorithm,
the resulting lines terminate before they can form a set of evenly spaced lines.
An example of this is shown in Fig.~\ref{fig:phase_space} where the truncated
\ac{UR} of the synthetic wavelength algorithm (square dots) is compared to the
ideal case (solid lines).
The irregular spacing of these ideal phase lines hint that the robustness of the
synthetic wavelength algorithm may vary with \ac{OPD}.

\begin{figure}[htb!]
  \centering
  \includegraphics[width=0.9\columnwidth]{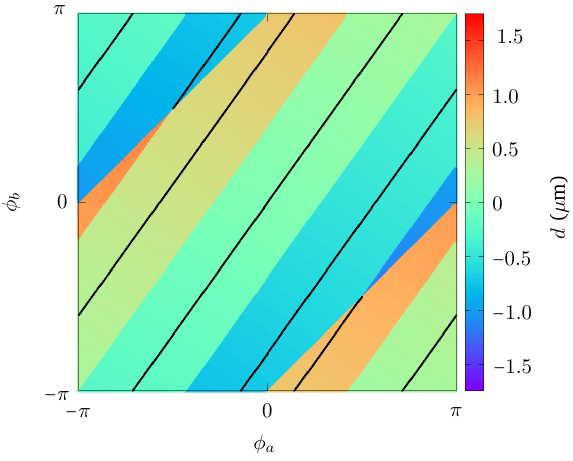}
  \caption[Synthetic Wavelength Alogrithm Output]{
    Calculated \ac{OPD} ($d$) for $\lambda_a = 700~\nm$ and $\lambda_b = 500~\nm$ using the
    synthetic wavelength algorithm. The solid line shows the expected phase
    values when measurement error is absent---note that the solid line does not
    evenly fill the phase space, as occurs in Fig.~\ref{fig:phase_space}.
    Similar to Fig.~\ref{fig:phase_space}, the color scale represents \ac{OPD}
    ($d$) and $d = 0~\um$ occurs at $(\phi_a, \phi_b) = (0, 0)$.
  }
  \label{fig:ps_partition}
\end{figure}

Note that any two-wavelength algorithm may be thought of as a function which
takes two phases as an input and returns an \ac{OPD}; \textit{i.e.}
$\ac{OPD} \coloneqq d(\phi_a, \phi_b)$.
With this in mind, and to further understand the behavior of the synthetic
wavelength algorithm, the algorithm can be applied to a raster scan of phase values,
$\phi_a$ and $\phi_b$, across the entire phase space, $[-\pi,\pi) \times [-\pi,\pi)$.
Fig.~\ref{fig:ps_partition} shows the \ac{OPD} as calculated using the synthetic
wavelength algorithm for $\lambda_a = 700~\nm$ and $\lambda_b = 500~\nm$ by
doing such a raster scan of phase space.  As we will show
below, the qualitative features seen in Fig.~\ref{fig:ps_partition} are common to
any choice of wavelength, even when the \ac{UR} of the synthetic wavelength
algorithm equals the maximum \ac{UR}.

The phase-mapping of \ac{OPD} shown in Fig.~\ref{fig:ps_partition} is
characterized by continuous bands of \ac{OPD} ($d$), where all points within
each continuous region correspond to a unique pair $(\bar{m}_a, \bar{m}_b)$.
The continuous bands of \ac{OPD} are bounded by discontinuities in $d$ that
represent sudden changes in the values of $\bar{m}_i$.
When the the \ac{OPD} is within the algorithm's \ac{UR}, and measurement error
is absent, the measured phases $(\phi_a, \phi_b)$ lie exactly on the ideal phase
(solid) lines shown in Fig.~\ref{fig:ps_partition}, and the values of
$\bar{m}_i$ are easily resolved.
When error is present, the measured phase is displaced from the ideal phase; as
long as the displacement is small enough to remain within the same continuous
band, the values of $\bar{m}_i$ are resolved correctly.
On the other hand, when the measured and ideal phases are separated by a
discontinuity boundary, the values of $\bar{m}_i$ are incorrectly resolved.

Near $d=0$, the discontinuities in the calculated \ac{OPD} are parallel to the ideal
phase lines, and the ideal phase lines are evenly spaced.  This appears to be a
common feature of all two-wavelength interferometry algorithms.  For the synthetic
wavelength algorithm, these discontinuities are located equidistant from adjacent
ideal phase lines.  In this region, the spacing between adjacent ideal phase
lines, $\Delta \phi_i$ is found by replacing $\abs{\ac{UR}_{\rm max}}$ with $\Lambda$ in
Eq.~\ref{eq:phase_offset}.  Thus, the values of $\bar{m}_i$ are determined
correctly when
\begin{equation}
  \label{eq:synthetic_constraint2}
  \lvert \delta \phi_\perp \rvert < \pi \left( \frac{\lambda_a - \lambda_b}{\sqrt{\lambda_a^2+\lambda_b^2}} \right).
\end{equation}
This corresponds to the well-known condition:
$\left \lvert \delta \bar{m}_i \right \rvert < 1/2$~\cite{Creath87,deGroot91b,Houairi_2009},
where $\delta \bar{m}$ is the error in $\bar{m}$ before rounding  to the nearest
integer.

For an \ac{OPD} in the region dominated by this condition, the synthetic wavelength
algorithm has robustness given by
\begin{equation}
  R = \left( \frac{1}{2\pi\sigma_a\sigma_b} \right)
  \int_{+\infty}^{-\infty}
  \int_{r_{\lambda}x-\frac{\Delta\phi_b}{2}}^{r_{\lambda}x+\frac{\Delta\phi_b}{2}}
  e^{-\frac{1}{2}\left( \frac{x^2}{\sigma_a^2} + \frac{y^2}{\sigma_b^2} \right)}
  \ud x \, \ud y \label{eq:robustness_messy}
  .
\end{equation}
When $\sigma_a = \sigma_b$, Eq.~\ref{eq:robustness_messy} is reduced to
\begin{equation}
  R = \text{erf} \left( \frac{\Delta \phi_a \Delta \phi_b}
                 {2 \sigma \sqrt{2 (\Delta \phi_a^2 + \Delta \phi_b^2)}} \right)
  \label{eq:hc_robustness}
  .
\end{equation}

Near the edges of the \ac{OPD} range, we see that the space between the ideal phase
lines and the discontinuities begins to shrink.  The change in the behavior of
the algorithm is related to a second condition
$\abs{\phi_{e,0} + \delta \phi_e} = \left \lvert \phi_{e,0} - \delta\phi_a + \delta\phi_b \right \rvert < \pi$
where $\phi_{e,0}$ is the effective phase in the absence of error.  Writing this
constraint in terms of $\delta \phi_\perp$ gives

\begin{figure}[htb!]
  \centering
  \includegraphics[width=\columnwidth]{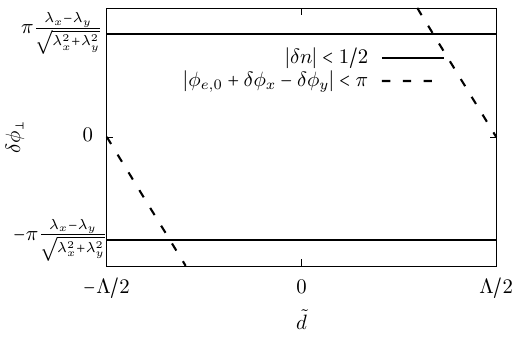}
  \caption[Synthetic Wavelength Algorithm Constraints]{
    Phase error constraints for the synthetic wavelength algorithm.  Note that
    these constraints agree with the discontinuities in Fig.~\ref{fig:ps_partition}.
    This can be visualized by ``stitching'' together the partitions of
    Fig.~\ref{fig:ps_partition} so that the ideal phase lines forms a
    continuous straight line with increasing \ac{OPD}.

  }
  \label{fig:synthetic_condition}
\end{figure}

\begin{IEEEeqnarray}{rCl}
  \label{eq:synthetic_constraint3}
  \left \lvert 2 \pi \frac{\tilde{d}}{\Lambda} - (\sin \theta + \cos \theta) \delta \phi_{\perp} \right \rvert & < & \pi
\end{IEEEeqnarray}

\noindent where $\theta$ is defined in Eq.~\ref{eq:transformation_angle}, and we've
used Eq.~\ref{eq:parallel_error_relation2} to rewrite the equation in terms of
the measured \ac{OPD} $\tilde{d}$, where $\tilde{d} = d_0 + \delta d$ and $d_0$
is the \ac{OPD} in the absence of any parallel phase error $\phi_\parallel$.  In
Fig.~\ref{fig:synthetic_condition}, these constraints are plotted across the
entire \ac{UR} for the synthetic wavelength algorithm when $\lambda_a = 700~\nm$
and $\lambda_b = 500~\nm$.

As shown in Fig.~\ref{fig:ps_partition}, the interaction of these constraints can effectively partition
the phase space in a non-uniform and uneven manner, making the correct
determination of the values of $\bar{m}_i$ much more challenging.
In Sec.~\ref{sec:hc_algorithm}, we discuss the robustness of the algorithms
presented in this paper in more detail (see Fig.~\ref{fig:robustness} in
Sec.~\ref{sec:hc_algorithm} where the robustness of the synthetic wavelength
algorithm over the \ac{UR} of the algorithm is shown in comparison to the
\ac{HC} algorithm).

\section{De~Groot Algorithm}\label{sec:de_groot}
De~Groot proposed an algorithm which extends the unambiguous range beyond that
of the synthetic wavelength algorithm.  To achieve this, de~Groot's algorithm
resolves the $\Lambda$ ambiguity which will arise when applying the synthetic
wavelength algorithm to range greater than $\pm \Lambda/2$~\cite{deGroot_1994}.
With this in mind, we note that the \ac{OPD} may be written as
\begin{equation}
  \label{eq:big_lambda_opd}
  d = \Lambda (\bar{M} + \dot{M}).
\end{equation}
We introduce a quantity which will prove important to the analysis of this
algorithm
\begin{equation}
  \label{eq:N_R}
  f_i = \abs{\frac{1}{\Lambda/\lambda_i - \round{\Lambda/\lambda_i}}}
  .
\end{equation}
It is straightforward to show that $f_a = f_b$ and $f > 2$ for all wavelengths.
The \ac{UR} of de~Groot's algorithm may be written as
$\ac{UR} = \pm \Lambda \, \round{f}/2$.

Following the procedure described in Sec.~\ref{sec:synthetic}, we apply the
de~Groot algorithm to a raster scan of phase values in the phase space.
When performing this analysis, we find that the algorithm exhibits qualitatively
different behavior depending on the difference between $f$ and the nearest
integer value, given by $f - \round{f}$.

\begin{figure}[htb!]
  \centering
  \hspace*{-0.8cm}
  \includegraphics[width=\columnwidth]{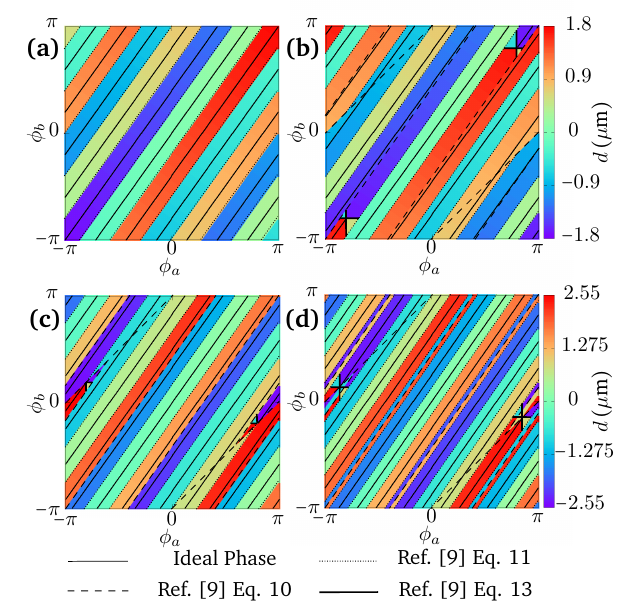}
  \caption[Wavelength Error]{
    Output of de~Groot's algorithm for (a) $f=2.00$, (b) $f=2.49$,
    (c) $f=2.51$, and (d) $f=3.49$.  Note that
    these values of $f$ are chosen to maximize the complex behavior of de~Groot's algorithm
    when $f - \round{f}$ is far from zero.  The wavelengths chosen are $500~\nm$
    and $700~\nm$, $500~\nm$ and $708.194~\nm$, $500~\nm$ and $708.472~\nm$, and
    $500~\nm$, $718.672~\nm$ respectively.  Each discontinuity is associated
    with an equation in de~Groot's paper~\cite{deGroot_1994}.  Note that some
    discontinuities are smaller than a wavelength, and, therefore, difficult to
    see.  The rounding function in Ref.~\cite{deGroot_1994}~Eq.~(12) does not
    create discontinuities in the algorithm's output.  The \ac{HC} algorithm
    produces the same output as de~Groot's algorithm for $\lambda_a=700~\nm$ and
    $\lambda_b=500~\nm$. Unlike de~Groot's algorith, the \ac{HC} algorithm
    always prodcues an even partitioning of the phase space.  As a result, the
    \ac{HC} algorithm will always produce an output similar to the figure in the
    top left.
  }
  \label{fig:de_groot_discontiuities}
\end{figure}

When $f - \round{f} = 0$, $f \in \mathbb{Z}$ and
$\ac{UR} = \pm \Lambda f/2$.  An example of the output of de~Groot's
algorithm when $f - \round{f} = 0$ is shown in
Fig.~\ref{fig:de_groot_discontiuities}a. In this scenario, de~Groot's algorithm
produces an even partitioning of the phase space.  As with the synthetic
wavelength algorithm, discontinuities in the algorithm's output are related to
constraints on the phase error.  For de~Groot's algorithm, this constraint is
\begin{IEEEeqnarray}{rCl}
  \label{eq:M-bar_constraint}
  \abs{\delta \bar{M}} & < & \frac{1}{2}
\end{IEEEeqnarray}
which may be rewritten in terms of $\delta \phi_{\perp}$ as
\begin{equation}
  \label{eq:degroot_constraint}
\lvert \delta \phi_\perp \rvert < \frac{\pi}{f} \left( \frac{\lambda_a - \lambda_b}{\sqrt{\lambda_a^2+\lambda_b^2}} \right)
  .
\end{equation}

The maximum allowed $\delta \phi_\perp$ for de~Groot's algorithm is smaller than
the maximum $\delta \phi_\perp$ allowed by the synthetic wavelength algorithm by a factor of
$f$.  However, de~Groot's algorithm extends the \ac{UR} of the synthetic wavelength
algorithm by a factor of $f$.  The inverse relationship between the maximum
allowed $\delta \phi_\perp$ and the \ac{UR} is a common feature to all
algorithms since extending \ac{UR} simply means that the bands of \ac{OPD}
continuity get more dense.  In terms of $U\!R/\lambda$, both the basic synthetic
wavelength algorithm and de~Groot's algorithm are
equally robust to phase error.  However, unlike the synthetic wavelength
algorithm, the robustness of de~Groot's algorithm does not deteriorate near the
edges of the \ac{UR}.

The output of de~Groot's algorithm becomes more complicated when
$f - \round{f} \ne 0$.  Eqs.~(10-13) from de~Groot's paper~\cite{deGroot_1994} feature
rounding functions that are used to extend the \ac{UR}.
Each rounding function is a source of discontinuity in the
algorithm's output.  In Fig.~\ref{fig:de_groot_discontiuities} we relate
the discontinuities in the algorithm's output to Eqs.~(10-13) from
Ref.~\cite{deGroot_1994} for different values of $f$.

While it is difficult make general statements regarding the robustness of de~Groot's
algorithm, a few observations can be made.  First, the robustness of de~Groot's
algorithm depends on \ac{OPD} when $f - \round{f} \ne 0$.  Second, when averaged over
the entire $\ac{UR}$, robustness decreases as $f - \round{f}$ moves further from
zero.

\section{Houairi \& Cassaing Algorithm}\label{sec:hc_algorithm}
An algorithm which achieves the maximum \ac{UR} for a given choice of
wavelengths was developed by
Houairi and Cassaing~\cite{Houairi_2009};
we refer
to this as the \ac{HC} algorithm.
The \ac{HC} algorithm utilizes the arithmetic properties of the fundamental
Diophantine equation relating phase measured by one wavelength to the phase
measured by the other wavelength, as described in Sec.~\ref{sec:max_ur}, to
disambiguate the phases to a greater extent than the typical synthetic
wavelength algorithm.
Unlike the other algorithms examined in this paper, the
spacing between discontinuities and the ideal phase lines for this algorithm remain constant over
the entire \ac{UR}.  The \ac{HC} algorithm will correctly resolve the values of
$\bar{m}_i$ as long as the phase errors satisfy the constraint~\cite{Houairi_2009}
\begin{equation}
  \label{eq:HC_constraint1}
  \left \lvert -p \delta \phi_a + q \delta \phi_b \right \rvert < \pi
  ,
\end{equation}
\noindent where $p$ and $q$ are as defined in Eq.~\ref{eq:UR_phase_condition1},
and also the constraint for the end of the \ac{UR} as shown in
Eq.~\ref{eq:HC_constraint3}.
The constraint in Eq.\ref{eq:HC_constraint1} may also be rewritten in terms of
$\delta \phi_{\perp}$ as
\begin{equation}
  \label{eq:HC_constraint2}
  \left \lvert \delta \phi_{\perp} \right \rvert < \frac{\pi}{\sqrt{p^2 + q^2}}
  .
\end{equation}

\begin{figure}[htb!]
  \centering
  \includegraphics[width=0.9\columnwidth]{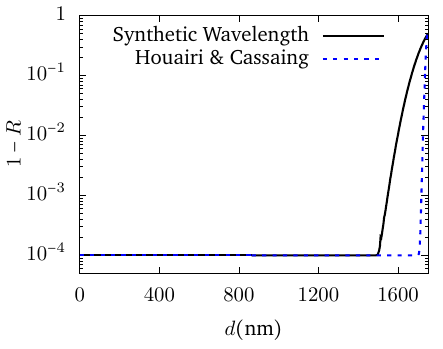}
  \caption[Algorithm Robustness]{
    Comparison of algorithm robustness.  \ac{HC} algorithm uses $\lambda_a = 700~\nm$,
    $\lambda_b = 500~\nm$ and $\sigma_a=\sigma_b = 0.0939$ radians.  Synthetic
    wavelength algorithm uses $\lambda_a = 619.824~\nm$, $\lambda_b = 526.572~\nm$,
    and $\sigma_a=\sigma_b = 0.0939$ radians; these
    wavelengths were chosen to achieve the same \ac{UR} and robustness of
    $0.9999$ at $d=0$.
  }
  \label{fig:robustness}
\end{figure}

\noindent Note that the HC algorithm is equivalent to de~Groot's algorithm when
$f - \round{f} = 0$.  The robustness of the \ac{HC} algorithm, calculated
following the description at the end of
Sec.~\ref{sec:phase_space_error}, results in Eq.~\ref{eq:robustness_messy}
or Eq.~\ref{eq:hc_robustness}, where $\Delta \phi_i$ are given by Eq.~\ref{eq:phase_offset}.  
In Fig.~\ref{fig:robustness} the robustness of the synthetic wavelength and
\ac{HC} algorithms are calculated and compared across the entire \ac{UR}.
Wavelengths $\lambda_a = 700~\nm$ and $\lambda_b = 500~\nm$ are used for the
\ac{HC} algorithm.  For the synthetic wavelength algorithm, wavelengths
$\lambda_a = 619.824~\nm$ and $\lambda_b=526.572~\nm$
are used to achieve a simliar \ac{UR} and robustness near $d=0$ as for the
\ac{HC} algorithm.  Phase errors of $\sigma_a=\sigma_b = 0.0939$
radians are used to calculate the robustness of both algorithms.  From
Fig.~\ref{fig:robustness},
we see the robustness of the synthetic wavelength algorithm deteriorates much
sooner than the \ac{HC} algorithm as $d \rightarrow \rm{UR}$.

\section{Wavelength Error}\label{sec:wavelength_error}
We have thus far neglected wavelength error in this analysis.  This
is justified as phase errors are typically on the order of $\pi / 100$ or worse.
Meanwhile, locking and measuring a wavelength with a fractional uncertainty of
$<10^{-4}$ is relatively simple using commercial servos and wavemeters.
Therefore, as we proceed, we will assume that errors in wavelength are small
compared to the phase errors.

\begin{figure}[htb!]
  \centering
  \includegraphics[width=0.85\columnwidth]{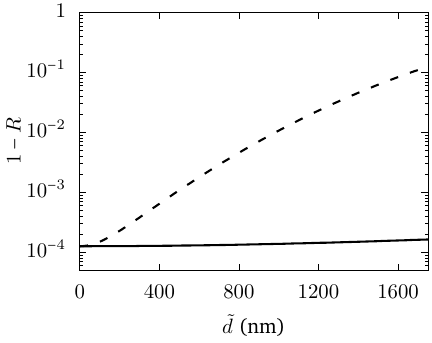}
  \caption[Wavelength Error Robustness]{
    Robustness of the \ac{HC} algorithm when for $\lambda_{0,a} = 700~\nm$,
    $\lambda_{0,b} = 500~\nm$ and $\sigma_a = \sigma_b = 0.0939$ with wavelength
    errors: $\delta \lambda_a = 0~\nm$ and $\delta \lambda_b = 1~\nm$ (solid);
    $\delta \lambda_a = 0~\nm$ and $\delta \lambda_b = 14~\nm$ (dashed).
  }
  \label{fig:wavelength_robustness}
\end{figure}

We denote the measured wavelength as $\tilde{\lambda}$, the actual wavelength
as $\lambda_0$ and the error in the measured wavelength as $\delta \lambda$, so
that $\tilde{\lambda} = \lambda_0 + \delta \lambda$.  The output of any
algorithm is determined by the user-specified measured wavelengths
$\tilde{\lambda}_i$.  Meanwhile, the ideal phase values, $\phi_0$, are
determined by the actual wavelength $\lambda_0$.  We may conceptually
understand the effects of wavelength error by plotting the
ideal phase values across phase space,
given by $\phi_{0,i} = 2 \pi \, d/\lambda_{0,i} \mod 2 \pi$,
against the \ac{OPD} ($d$) from the \ac{HC} algorithm using $\tilde{\lambda}_i$.
\begin{figure}[htb!]
  \centering
  \includegraphics[width=0.9\columnwidth]{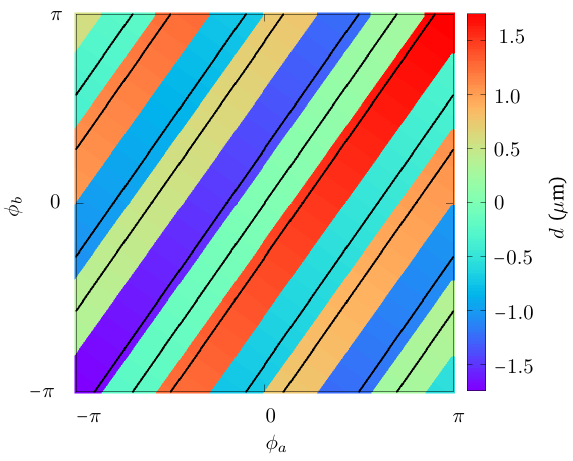}
  \caption[Wavelength Error]{
    Output of the \ac{HC} algorithm for $\lambda_{0,a} = 700~\nm$ and
    $\lambda_{0,b} = 500~\nm$ with ideal phase values for
    $\tilde{\lambda}_a = 714~\nm$ and $\tilde{\lambda}_b = 500~\nm$.
  }
  \label{fig:wavelength_error}
\end{figure}
This phase space representation presented in Fig.~\ref{fig:wavelength_error}
shows that the slope of the
ideal phase lines no longer match the slope of the discontinuities in the
algorithm's output.  As a result, ideal phase values move closer to one
discontinuity and further from the other.  This can be thought of as a
systematic contribution to $\delta \phi_\perp$.  The relationship between
$\delta \lambda_i$ and the contribution to $\delta \phi_\perp$ caused by
$\delta \lambda_i$, which we denote as $\delta \phi_{\lambda,\perp}$, is given by
\begin{equation}
  \label{eq:wavelength_phase_error}
  \delta \phi_{\lambda,\perp} = 2 \pi \left( \frac{\delta r_\lambda}{r_\lambda} \right) \frac{d}{\sqrt{\lambda_a^2+\lambda_b^2}}
\end{equation}
where $r_\lambda = \lambda_a/\lambda_b$, so that $\delta r_\lambda$ is given by
\begin{equation}
  \label{eq:delta_r}
  \delta r_\lambda = \frac{\lambda_b \delta \lambda_a - \lambda_a \delta \lambda_b}{\lambda_b^2}
  .
\end{equation}
When the phase uncertainty is the same for both phases and considering both
phase and wavelength errors, the robustness of the
\ac{HC} algorithm is then given by
\begin{equation}
  \label{eq:robustness}
  \begin{split}
  R \approx \frac{1}{2} \vast(
      &\mbox{erf} \left(
          \frac{\pi \, \lambda_x \, \lambda_y}
               {\sigma \, \abs{\ac{UR}} \sqrt{2(\lambda_x^2+\lambda_y^2)}}
        - \frac{\delta \phi_{\lambda,\perp}}{\sqrt{2} \sigma}
      \right) \\
    - &\mbox{erf} \left(
        - \frac{\pi \, \lambda_x \, \lambda_y}
               {\sigma \, \abs{\ac{UR}} \sqrt{2(\lambda_x^2+\lambda_y^2)}}
        - \frac{\delta \phi_{\lambda,\perp}}{\sqrt{2} \sigma}
      \right)
  \vast)
  \end{split}
\end{equation}
Fig.\ref{fig:wavelength_robustness} demonstrates the robustness $R$ from
Eq.~\ref{eq:robustness} resulting from either a small or a large error in one of
the two wavelengths used for the \ac{HC} algorithm.

\section{Conclusions}\label{sec:conclusion}

In this paper we introduced the idea of the phase space as a tool for analyzing
the behavior of multi-wavelength interferometry algorithms.  We show that the
component of the phase error perpendicular to the ideal phase lines is solely
responsible for algorithm success.  When the phase uncertainty is the same for
both wavelengths, we find that the component of the phase error parallel to the
ideal phase lines is responsible for error in the calculated \ac{OPD}.

We note that the robustness of an algorithm which does not achieve the maximum
\ac{UR} will likely depend on the \ac{OPD}.  In particular, we show that the robustness of
the synthetic wavelength algorithm decreases when the \ac{OPD} is within $\lambda/2$
of $\pm \Lambda/2$.  We show that the drop in robustness is associated with an
overlooked constraint of the synthetic wavelength algorithm.  We show that
robustness of de~Groot's algorithm depends on wavelength choice and \ac{OPD} in a
non-trivial way.  Finally we show that the \ac{HC} algorithm results in an optimal
partitioning of the phase space.  This even partitioning results in uniform
robustness across the entire \ac{UR}.  We examined the effect of wavelength error on
the robustness of the \ac{HC} algorithm and found that wavelength errors result in
robustness which decreases as $|d|$ increases.

\textbf{Funding} This work was funded by the Air Force Office of Scientific Research under lab
task 22RVCOR017.

\textbf{Disclaimer}
The views expressed are those of the authors and do not necessarily reflect the
official policy or position of the Department of the Air Force, the Department
of the Defense, or the U.S. Government.

\bibliographystyle{apsrev4-1}
\bibliography{profilometry}

\begin{thebibliography}{16}%
\makeatletter
\providecommand \@ifxundefined [1]{%
 \@ifx{#1\undefined}
}%
\providecommand \@ifnum [1]{%
 \ifnum #1\expandafter \@firstoftwo
 \else \expandafter \@secondoftwo
 \fi
}%
\providecommand \@ifx [1]{%
 \ifx #1\expandafter \@firstoftwo
 \else \expandafter \@secondoftwo
 \fi
}%
\providecommand \natexlab [1]{#1}%
\providecommand \enquote  [1]{``#1''}%
\providecommand \bibnamefont  [1]{#1}%
\providecommand \bibfnamefont [1]{#1}%
\providecommand \citenamefont [1]{#1}%
\providecommand \href@noop [0]{\@secondoftwo}%
\providecommand \href [0]{\begingroup \@sanitize@url \@href}%
\providecommand \@href[1]{\@@startlink{#1}\@@href}%
\providecommand \@@href[1]{\endgroup#1\@@endlink}%
\providecommand \@sanitize@url [0]{\catcode `\\12\catcode `\$12\catcode
  `\&12\catcode `\#12\catcode `\^12\catcode `\_12\catcode `\%12\relax}%
\providecommand \@@startlink[1]{}%
\providecommand \@@endlink[0]{}%
\providecommand \url  [0]{\begingroup\@sanitize@url \@url }%
\providecommand \@url [1]{\endgroup\@href {#1}{\urlprefix }}%
\providecommand \urlprefix  [0]{URL }%
\providecommand \Eprint [0]{\href }%
\providecommand \doibase [0]{http://dx.doi.org/}%
\providecommand \selectlanguage [0]{\@gobble}%
\providecommand \bibinfo  [0]{\@secondoftwo}%
\providecommand \bibfield  [0]{\@secondoftwo}%
\providecommand \translation [1]{[#1]}%
\providecommand \BibitemOpen [0]{}%
\providecommand \bibitemStop [0]{}%
\providecommand \bibitemNoStop [0]{.\EOS\space}%
\providecommand \EOS [0]{\spacefactor3000\relax}%
\providecommand \BibitemShut  [1]{\csname bibitem#1\endcsname}%
\let\auto@bib@innerbib\@empty
\bibitem [{\citenamefont {de~Groot}(1994)}]{deGroot_1994}%
  \BibitemOpen
  \bibfield  {author} {\bibinfo {author} {\bibfnamefont {P.~J.}\ \bibnamefont
  {de~Groot}},\ }\href {\doibase 10.1364/AO.33.005948} {\bibfield  {journal}
  {\bibinfo  {journal} {Appl. Opt.}\ }\textbf {\bibinfo {volume} {33}},\
  \bibinfo {pages} {5948} (\bibinfo {year} {1994})}\BibitemShut {NoStop}%
\bibitem [{\citenamefont {Houairi}\ and\ \citenamefont
  {Cassaing}(2009)}]{Houairi_2009}%
  \BibitemOpen
  \bibfield  {author} {\bibinfo {author} {\bibfnamefont {K.}~\bibnamefont
  {Houairi}}\ and\ \bibinfo {author} {\bibfnamefont {F.}~\bibnamefont
  {Cassaing}},\ }\href {\doibase 10.1364/JOSAA.26.002503} {\bibfield  {journal}
  {\bibinfo  {journal} {J. Opt. Soc. Am. A}\ }\textbf {\bibinfo {volume}
  {26}},\ \bibinfo {pages} {2503} (\bibinfo {year} {2009})}\BibitemShut
  {NoStop}%
\bibitem [{\citenamefont {Chiaradia}\ \emph {et~al.}(1996)\citenamefont
  {Chiaradia}, \citenamefont {Guerriero}, \citenamefont {Pasquariello},
  \citenamefont {Refice},\ and\ \citenamefont {Veneziani}}]{Chiaradia_1996}%
  \BibitemOpen
  \bibfield  {author} {\bibinfo {author} {\bibfnamefont {M.}~\bibnamefont
  {Chiaradia}}, \bibinfo {author} {\bibfnamefont {L.}~\bibnamefont
  {Guerriero}}, \bibinfo {author} {\bibfnamefont {G.}~\bibnamefont
  {Pasquariello}}, \bibinfo {author} {\bibfnamefont {A.}~\bibnamefont
  {Refice}}, \ and\ \bibinfo {author} {\bibfnamefont {N.}~\bibnamefont
  {Veneziani}},\ }in\ \href {\doibase 10.1109/IGARSS.1996.516888} {\emph
  {\bibinfo {booktitle} {IGARSS '96. 1996 International Geoscience and Remote
  Sensing Symposium}}},\ Vol.~\bibinfo {volume} {4}\ (\bibinfo {year} {1996})\
  pp.\ \bibinfo {pages} {2060--2062 vol.4}\BibitemShut {NoStop}%
\bibitem [{\citenamefont {Veneziani}\ \emph {et~al.}(2003)\citenamefont
  {Veneziani}, \citenamefont {Bovenga},\ and\ \citenamefont
  {Refice}}]{Veneziani_2003}%
  \BibitemOpen
  \bibfield  {author} {\bibinfo {author} {\bibfnamefont {N.}~\bibnamefont
  {Veneziani}}, \bibinfo {author} {\bibfnamefont {F.}~\bibnamefont {Bovenga}},
  \ and\ \bibinfo {author} {\bibfnamefont {A.}~\bibnamefont {Refice}},\ }\href
  {\doibase 10.1007/978-1-4757-6342-3_9} {\bibfield  {journal} {\bibinfo
  {journal} {Multidimensional Systems and Signal Processing}\ }\textbf
  {\bibinfo {volume} {14}},\ \bibinfo {pages} {183} (\bibinfo {year}
  {2003})}\BibitemShut {NoStop}%
\bibitem [{\citenamefont {Xu}\ \emph {et~al.}(2016)\citenamefont {Xu},
  \citenamefont {An}, \citenamefont {Huang},\ and\ \citenamefont
  {Wang}}]{Xu_2016}%
  \BibitemOpen
  \bibfield  {author} {\bibinfo {author} {\bibfnamefont {J.}~\bibnamefont
  {Xu}}, \bibinfo {author} {\bibfnamefont {D.}~\bibnamefont {An}}, \bibinfo
  {author} {\bibfnamefont {X.}~\bibnamefont {Huang}}, \ and\ \bibinfo {author}
  {\bibfnamefont {G.}~\bibnamefont {Wang}},\ }\href {\doibase
  https://doi.org/10.1049/iet-rsn.2015.0328} {\bibfield  {journal} {\bibinfo
  {journal} {IET Radar, Sonar \& Navigation}\ }\textbf {\bibinfo {volume}
  {10}},\ \bibinfo {pages} {426} (\bibinfo {year} {2016})}\BibitemShut
  {NoStop}%
\bibitem [{\citenamefont {Siebert}\ \emph {et~al.}(2004)\citenamefont
  {Siebert}, \citenamefont {Splitthof},\ and\ \citenamefont
  {Ettemeyer}}]{Siebert_2004}%
  \BibitemOpen
  \bibfield  {author} {\bibinfo {author} {\bibfnamefont {T.}~\bibnamefont
  {Siebert}}, \bibinfo {author} {\bibfnamefont {K.}~\bibnamefont {Splitthof}},
  \ and\ \bibinfo {author} {\bibfnamefont {A.}~\bibnamefont {Ettemeyer}},\
  }\href {\doibase 10.1166/jhs.2004.005} {\bibfield  {journal} {\bibinfo
  {journal} {Journal of Holography and Speckle}\ }\textbf {\bibinfo {volume}
  {1}},\ \bibinfo {pages} {32} (\bibinfo {year} {2004})}\BibitemShut {NoStop}%
\bibitem [{\citenamefont {Yankelev}\ \emph {et~al.}(2020)\citenamefont
  {Yankelev}, \citenamefont {Avinadav}, \citenamefont {Davidson},\ and\
  \citenamefont {Firstenberg}}]{Yankelev_2020}%
  \BibitemOpen
  \bibfield  {author} {\bibinfo {author} {\bibfnamefont {D.}~\bibnamefont
  {Yankelev}}, \bibinfo {author} {\bibfnamefont {C.}~\bibnamefont {Avinadav}},
  \bibinfo {author} {\bibfnamefont {N.}~\bibnamefont {Davidson}}, \ and\
  \bibinfo {author} {\bibfnamefont {O.}~\bibnamefont {Firstenberg}},\ }\href
  {\doibase 10.1126/sciadv.abd0650} {\bibfield  {journal} {\bibinfo  {journal}
  {Science Advances}\ }\textbf {\bibinfo {volume} {6}},\ \bibinfo {pages}
  {eabd0650} (\bibinfo {year} {2020})}\BibitemShut {NoStop}%
\bibitem [{\citenamefont {Wyant}\ \emph {et~al.}(1984)\citenamefont {Wyant},
  \citenamefont {Oreb},\ and\ \citenamefont {Hariharan}}]{Wyant_1984}%
  \BibitemOpen
  \bibfield  {author} {\bibinfo {author} {\bibfnamefont {J.~C.}\ \bibnamefont
  {Wyant}}, \bibinfo {author} {\bibfnamefont {B.~F.}\ \bibnamefont {Oreb}}, \
  and\ \bibinfo {author} {\bibfnamefont {P.}~\bibnamefont {Hariharan}},\ }\href
  {\doibase 10.1364/AO.23.004020} {\bibfield  {journal} {\bibinfo  {journal}
  {Appl. Opt.}\ }\textbf {\bibinfo {volume} {23}},\ \bibinfo {pages} {4020}
  (\bibinfo {year} {1984})}\BibitemShut {NoStop}%
\bibitem [{\citenamefont {Ribbens}(1974)}]{Ribbens_1974}%
  \BibitemOpen
  \bibfield  {author} {\bibinfo {author} {\bibfnamefont {W.~B.}\ \bibnamefont
  {Ribbens}},\ }\href {\doibase 10.1364/AO.13.001085} {\bibfield  {journal}
  {\bibinfo  {journal} {Appl. Opt.}\ }\textbf {\bibinfo {volume} {13}},\
  \bibinfo {pages} {1085} (\bibinfo {year} {1974})}\BibitemShut {NoStop}%
\bibitem [{\citenamefont {Polhemus}(1973)}]{Polhemus_1973}%
  \BibitemOpen
  \bibfield  {author} {\bibinfo {author} {\bibfnamefont {C.}~\bibnamefont
  {Polhemus}},\ }\href {\doibase 10.1364/AO.12.002071} {\bibfield  {journal}
  {\bibinfo  {journal} {Appl. Opt.}\ }\textbf {\bibinfo {volume} {12 9}},\
  \bibinfo {pages} {2071} (\bibinfo {year} {1973})}\BibitemShut {NoStop}%
\bibitem [{\citenamefont {Falaggis}\ \emph {et~al.}(2008)\citenamefont
  {Falaggis}, \citenamefont {Towers},\ and\ \citenamefont
  {Towers}}]{Falaggis_2008}%
  \BibitemOpen
  \bibfield  {author} {\bibinfo {author} {\bibfnamefont {K.}~\bibnamefont
  {Falaggis}}, \bibinfo {author} {\bibfnamefont {D.~P.}\ \bibnamefont
  {Towers}}, \ and\ \bibinfo {author} {\bibfnamefont {C.~E.}\ \bibnamefont
  {Towers}},\ }in\ \href {\doibase 10.1117/12.795293} {\emph {\bibinfo
  {booktitle} {Interferometry XIV: Techniques and Analysis}}},\ Vol.\ \bibinfo
  {volume} {7063},\ \bibinfo {editor} {edited by\ \bibinfo {editor}
  {\bibfnamefont {J.}~\bibnamefont {Schmit}}, \bibinfo {editor} {\bibfnamefont
  {K.}~\bibnamefont {Creath}}, \ and\ \bibinfo {editor} {\bibfnamefont {C.~E.}\
  \bibnamefont {Towers}}},\ \bibinfo {organization} {International Society for
  Optics and Photonics}\ (\bibinfo  {publisher} {SPIE},\ \bibinfo {year}
  {2008})\ pp.\ \bibinfo {pages} {318 -- 325}\BibitemShut {NoStop}%
\bibitem [{\citenamefont {Lofdahl}\ and\ \citenamefont
  {Eriksson}(2001)}]{Lofdahl_2001}%
  \BibitemOpen
  \bibfield  {author} {\bibinfo {author} {\bibfnamefont {M.~G.}\ \bibnamefont
  {Lofdahl}}\ and\ \bibinfo {author} {\bibfnamefont {H.}~\bibnamefont
  {Eriksson}},\ }\href {\doibase 10.1117/1.1365936} {\bibfield  {journal}
  {\bibinfo  {journal} {Opt. Eng.}\ }\textbf {\bibinfo {volume} {40}},\
  \bibinfo {pages} {984 } (\bibinfo {year} {2001})}\BibitemShut {NoStop}%
\bibitem [{\citenamefont {van Brug}\ and\ \citenamefont
  {Klaver}(1998)}]{Brug_1998}%
  \BibitemOpen
  \bibfield  {author} {\bibinfo {author} {\bibfnamefont {H.}~\bibnamefont {van
  Brug}}\ and\ \bibinfo {author} {\bibfnamefont {R.~G.}\ \bibnamefont
  {Klaver}},\ }\href {\doibase 10.1088/0963-9659/7/6/023} {\bibfield  {journal}
  {\bibinfo  {journal} {Pure \& Appl. Opt.: J. Euro. Opt. Soc. A}\ }\textbf
  {\bibinfo {volume} {7}},\ \bibinfo {pages} {1465} (\bibinfo {year}
  {1998})}\BibitemShut {NoStop}%
\bibitem [{\citenamefont {Wyant}(1971)}]{Wyant71}%
  \BibitemOpen
  \bibfield  {author} {\bibinfo {author} {\bibfnamefont {J.~C.}\ \bibnamefont
  {Wyant}},\ }\href {http://dx.doi.org/10.1364/AO.10.002113} {\bibfield
  {journal} {\bibinfo  {journal} {Appl. Opt.}\ }\textbf {\bibinfo {volume}
  {10}},\ \bibinfo {pages} {2113 } (\bibinfo {year} {Sept. 1971})}\BibitemShut
  {NoStop}%
\bibitem [{\citenamefont {Creath}(1987)}]{Creath87}%
  \BibitemOpen
  \bibfield  {author} {\bibinfo {author} {\bibfnamefont {K.}~\bibnamefont
  {Creath}},\ }\href {\doibase 10.1364/AO.26.002810} {\bibfield  {journal}
  {\bibinfo  {journal} {Appl. Opt.}\ }\textbf {\bibinfo {volume} {26}},\
  \bibinfo {pages} {2810} (\bibinfo {year} {1987})}\BibitemShut {NoStop}%
\bibitem [{\citenamefont {de~Groot}\ and\ \citenamefont
  {Kishner}(1991)}]{deGroot91b}%
  \BibitemOpen
  \bibfield  {author} {\bibinfo {author} {\bibfnamefont {P.}~\bibnamefont
  {de~Groot}}\ and\ \bibinfo {author} {\bibfnamefont {S.}~\bibnamefont
  {Kishner}},\ }\href {\doibase 10.1364/AO.30.004026} {\bibfield  {journal}
  {\bibinfo  {journal} {Appl. Opt.}\ }\textbf {\bibinfo {volume} {30}}
  (\bibinfo {year} {1991}),\ 10.1364/AO.30.004026}\BibitemShut {NoStop}%
\end{thebibliography}%
\end{document}